\documentclass{article}
\usepackage[paper=letterpaper,margin=2cm]{geometry}
\usepackage{soul, color, xcolor}
\usepackage{graphicx}
\usepackage{amsmath}
\usepackage{amssymb}
\usepackage{braket}
\usepackage{bm}
\usepackage{amsfonts}
\usepackage{cite}
\usepackage[ruled,boxed,linesnumbered]{algorithm2e}
\usepackage{tcolorbox}
\usepackage[colorlinks=true,citecolor=blue]{hyperref}
\usepackage{booktabs}
\usepackage{authblk}
\date{}
\title{\textbf{Experimental Frequency-Comb-Based Mode-Pairing Quantum Key Distribution \\Beyond the Rate-Loss Limit}}

\author{\begin{minipage}{0.92\textwidth}
    \centering
    \small
    Yi-Fei Lu$^1$, Yan-Yang Zhou$^{1,\dagger}$, Yu-Yao Guo$^2$, Xin-Hang Li$^2$, Xiao-Lei Jiang$^1$, Yang Wang$^1$, 
    
    Yu Zhou$^1$, Jia-Ji Li$^1$, Chun Zhou$^1$, Hong-Wei Li$^1$, Lin-Jie Zhou$^2$, Wan-Su Bao$^{1,\ddagger}$ \\
    $^1$Henan Key Laboratory of Quantum Information and Cryptography, IEU, Zhengzhou 450001, China\\
    $^2$State Key Laboratory of Advanced Optical Communication Systems and Networks, \\Shanghai Key Lab of Navigation and Location Services, Department of Electronic Engineering, \\Shanghai Jiao Tong University, Shanghai 200240, China\\
    $^\dagger$zyy@qiclab.cn, $^\ddagger$bws@qiclab.cn
\end{minipage}}

\begin{document}
\maketitle

{\centering\section*{Abstract}}

Mode-pairing quantum key distribution (MP-QKD) offers significant potential for long-distance secure communication, benefiting from its quadratic scaling capacity and phase compensation-free characteristic. However, MP-QKD still requires stringent wavelength consistency between remote lasers, which is impractical with commercial lasers without locking. In this work, we develop a simple optical-frequency-comb-based MP-QKD system to simultaneously establish coherence and distribute keys with free-running commercial lasers. We implement the experiment over standard single-mode fiber with a loss coefficient of 0.2 dB/km. The system achieves finite-size secret key rates (SKRs) of 561.26, 113.59, and 10.20 bits per second over 303.37, 354.62, and 404.25 km, which are 1.0267, 2.5230, and 2.1033 times of the rate-loss limit. Besides, the SKR over 202.31 km is 8695.34 bits per second, which is sufficient to support practical cryptographic applications. These results validate the practicality and robustness of our method, offering a simplified yet high-performance framework for long-distance secure communication.

\clearpage

\section{Introduction}

Quantum key distribution (QKD) \cite{bennett2014RN153,lo1999RN138,xu2020RN130,pirandola2020RN485,portmann2022RN813} represents a revolutionary approach in secure communication, leveraging the laws of quantum mechanics to share secret keys between remote two parties (named Alice and Bob). Numerous revolutionary advancements have been present in both the theory \cite{grosshans2002RN526,long2002RN1066,hwang2003RN202,lo2005RN81,wang2005RN185,barrett2005RN516,lo2012RN72,braunstein2012RN581,islam2017RN762} and experimentation \cite{yin2016RN288,yuan2018RN513,boaron2018RN286,wei2020RN1030,chen2021RN519,fadri2023RN1023,li2023RN987}. However, the achievable secret key rate (SKR) in point-to-point QKD systems is fundamentally limited by channel loss \cite{takeoka2014RN231, pirandola2017RN103}, which significantly restricts long-distance applications. Even if long-distance QKD is realized, it comes at the price of an impractically low key rate.

To overcome this fundamental rate-distance limit \cite{takeoka2014RN231, pirandola2017RN103}, twin-field (TF) QKD \cite{lucamarini2018RN45,ma2018RN56,wang2018RN22,curty2019RN57,cui2019RN41,maeda2019RN507} was proposed based on the single-photon interference. In TF-QKD, the physical carriers are coherent states, and the information is encoded in the global phase, which contributed to high key rate and long-distance capabilities \cite{fang2020RN55,chen2021RN586,pittaluga2021RN482,wang2022RN585,li2023RN979,liu2023RN975,zhou2023RN949}. Meanwhile, the first-order interference characteristics demand high wavelength consistency between Alice and Bob's lasers, as well as rapid compensation for phase drift throughout the entire channels. Though these challenges have been overcome with several technologies, e.g., injection locking, optical phase-locked loop, time-frequency dissemination, and dual-band stabilization, they introduce significant complexity and may hinder practical application. To reduce experimental complexity while maintaining high performance, mode-pairing (MP) QKD  (also known as asynchronous measurement-device-independent (MDI) QKD) has been proposed \cite{zeng2022RN816,xie2022RN724}. MP-QKD uses the time-bin basis as the encoding basis and the phase basis as the test basis, which can be regarded as an enhanced version of the traditional MDI-QKD \cite{lo2012RN72,ma2012RN73}. MP-QKD can comparatively decouple the paired bins and post-determine their pairing based on Charlie's announced results. This enables a scaling of the SKR as $O(\sqrt{\eta})$ \cite{zeng2022RN816,xie2022RN724,lu2025RN1293}, where $\eta$ denotes the transmission efficiency between Alice and Bob. Besides, as the encoding is based on the relationship between two time bins, the rapid phase drift in channels is compensated automatically. Hence MP-QKD eliminates the need for rapid phase compensation and is advantageous for long-distance secure communication.

Nevertheless, to obtain high performance, MP-QKD still demands a high requirement of wavelength consistency between remote lasers. The error rate in the test basis is mainly determined by the residual phase $2\pi \Delta f \Delta t$, where $\Delta f$ is the wavelength difference and $\Delta t$ is the time delay in paired bins. There exists a trade-off between the pairing efficiency and the error rate. A larger pairing interval leads to higher pairing efficiency, but it may result in an increased error rate in the test basis. Consequently, overcoming the issues of laser frequency consistency and stabilization remains a pivotal challenge for both high SKR and long-distance QKD.  Additionally, the phase drift in the channel can also introduce errors, but its impact is mainly observed under unstable environmental conditions. The first MP-QKD demonstration \cite{zhu2023RN928} utilized a time-division approach to send strong reference light for real-time monitoring of the frequency difference between remote lasers based on the maximum-likelihood estimation method. It demonstrated a square-root scaling SKR with a laser linewidth of 2 kHz. To surpass the fundamental rate-distance limit, it employed two ultra-stable lasers, which use the Pound-Drever-Hall technique for locking to a reference cavity, and obtain a short-term linewidth of 1 Hz. Based on these conditions, it achieved an SKR that was 4.08 times the absolute rate limit over 508 km ultra-low-loss fiber \cite{zhou2023RN978}. Besides, to stabilize the lasers, it references the lasers to an absolute frequency standard with the acetylene $(^{13}{\text C}_2{\text H}_2){\text P}_{16}(\nu_1+\nu_3)$ transition line at 1542.384 nm \cite{ge2025RN1181}. Then it also demonstrates a breakthrough of the fundamental rate-distance limit \cite{ge2025RN1181}. Therefore, it currently remains a challenge to achieve a practical and efficient MP-QKD system surpassing the rate-loss limit based on commercial lasers without local locking.

In this work, we employ two free-running commercial lasers to implement MP-QKD and demonstrate its performance surpassing the rate-loss limit.  To overcome the limitations of wavelength inconsistency and instability, we generate the optical frequency combs at each sender and develop a sliding frequency referencing method. The coherence between different sidebands allows us to perform key distribution and frequency referencing simultaneously, while also allowing the transmission in the same channel. Specifically, we use the H34 channel sideband for key distribution and the C34 channel sideband to establish coherence. This also enables the use of strong reference light to guarantee adequate reference counts, along with precise referencing based on light from the same time interval. Besides, we note the sliding frequency referencing method will not compromise efficiency as it allows full-time key generation and MP-QKD inherently requires phase randomization of the signal pulses. In the experiments, we employ the standard single-mode fiber as the channel with a typical loss coefficient of 0.2 dB/km. We achieve a finite-size SKR of 8.7 kbit per second over 202.31 km. And we achieve the finite-size SKRs of 561, 114, and 10 bits per second at 303.37, 354.62, and 404.25 km, which surpass the absolute rate-loss limit by a factor of 1.03, 2.52, and 2.10, respectively.

\section{Protocol}

In this section, we present the vacuum + weak decoy-state MP-QKD protocol \cite{zeng2022RN816,xie2022RN724}.

Step 1 (State preparation). In the $j$-th round, Alice chooses a intensity $\tau_j^a \in \{\mu_a, \nu_a, o_a\}$ with probability $p_{\tau_j^a}$ and a random phase $\theta_j^a$. Here, the intensities satisfy $0 = o_a < \nu_a < \mu_a$. Then she prepares the coherent state $\ket{e^{i\theta_j^a} \sqrt{\tau_j^a}}$ and sends it to Charlie. Bob independently performs parameter selection and state preparation in the same manner.

Step 2 (Measurement). Charlie performs the interference measurement with a 50:50 beam splitter (BS) along with two single-photon detectors (SPDs). Then Charlie announces the measurement results, denoted as $L_j,R_j \in \mathbb{Z}_2$, of the two SPDs, where '1' and '0' indicate the SPD click or not. The honesty of Charlie is not necessary and the deception will affect the secret key rate (SKR) but will not cause security issues.

Step 3 (Mode pairing). After repeating the above steps $N$ times, Alice and Bob perform the mode pairing and obtain the pairs according to the pairing strategy in Algorithm \ref{pairing_strategy}. The basic idea is to first filter out some invalid rounds, then pair two adjacent valid rounds, while ensuring that the pairing is constrained within a maximum pairing interval $L_{\max}$.

Step 4 (Basis sifting). Alice assigns the pairs (indexed by $j$ and $k$) as Z or X basis according to the intensities of paired rounds. Specifically, a pair is assigned as Z basis if $\{\tau_j^a,\tau_k^a\} = \{\mu_a, o_a\}$, and as X basis if $\tau_j^a=\tau_k^a = \nu_a$. Similarly, Bob assigns the basis locally.

Step 5 (Key mapping). In Z basis, Alice extracts the raw key bits as 0 or 1 depending on whether $\tau_j^a = \mu_a$ or $\tau_k^a = \mu_a$, while Bob conversely extracts the raw key bits as 1 or 0 if $\tau_j^a = \mu_a$ or $\tau_k^a = \mu_a$. When both Alice and Bob are X basis, Alice and Bob announce the phase differences $\delta_{jk}^a = (\theta_j^a - \theta_k^a) \mod 2\pi$ and $\delta_{jk}^b = (\theta_j^b - \theta_k^b) \mod 2\pi$. They sift those pairs satisfying $|\delta_{jk}^a - \delta_{jk}^b| \leq 2\pi/M$ or $||\delta_{jk}^a - \delta_{jk}^b| - \pi| \leq 2\pi/M$ to estimate the information leakage. Here, $M$ is the number of the phase slices and is often set as 16 to balance the number of pairs and the error rate. The error pairs are defined as two cases: $L_j = L_k$ when $||\delta_{jk}^a - \delta_{jk}^a| - \pi| \leq 2\pi/M$, $L_j = R_k$ when $|\delta_{jk}^a - \delta_{jk}^a| \leq 2\pi/M$.

Step 6 (Parameter estimation). By employing the decoy-state method, they estimate the lower bound of the number of bits $\underline{n}_{11}^z$ and the upper bound of the phase error rate $\overline{e}_{11}^{\text{ph}}$ when both Alice and Bob send the single-photon states in Z basis.

Step 7 (Parameter estimation and postprocessing). Alice and Bob perform the error correction and privacy amplification procedures to distill the final key bits. The SKR in the finite case can be shown as \cite{zeng2022RN816,xie2022RN724,zhou2023RN949}
\begin{equation}
    R = \frac{1}{N} \Big\{ \underline{n}_{11}^z [1 - H_2 (\overline{e}_{11}^{\text{ph}})] - \lambda_{\text{EC}} -\log_2\frac{2}{\varepsilon_{\text{cor}}} - 2 \log_2 \frac{2}{\varepsilon^\prime \hat{\varepsilon}} - 2 \log_2 \frac{1}{\varepsilon_{\text{PA}}} \Big\},
\end{equation}
where $H_2 (x) = x \log_2(x) - (1-x) \log_2(1-x)$ is the binary Shannon entropy function, $\lambda_{\text{EC}}$ is the information revealed in the error correction step, and $\varepsilon_{\text{cor}}$, $\varepsilon^\prime$, $\hat{\varepsilon}$, and $\varepsilon_{\text{PA}}$ are the security coefficients regarding the correctness and secrecy.

\section{Experimental setup}

\begin{figure}
    \centering
    \includegraphics[width=0.96\textwidth]{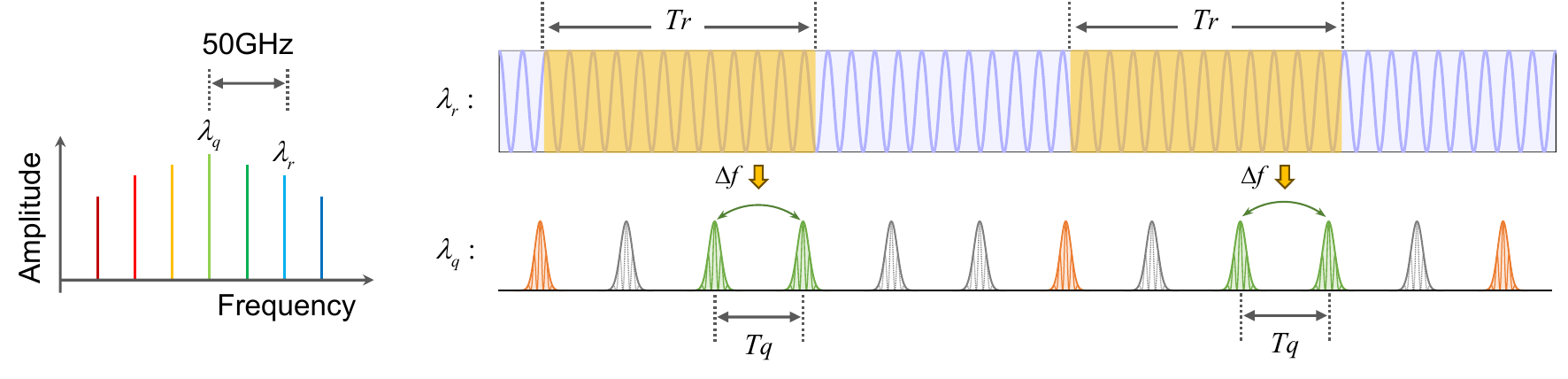}
    \caption{Sliding frequency referencing method. The source is the optic frequency comb. Two sidebands $\lambda_q$ and $\lambda_r$ are selected for quantum encoding and establishment of coherence, respectively. The quantum light is chopped into a pulse train with period $T_q$. The frequency reference light is attenuated into a proper intensity. The right illustrates the compensation method when the maximum pairing interval is limited to 1. The gray pulses represent invalid detections, the orange pulses represent valid detections but with failed pairing, and the green pulses represent the rounds of successful pairing. When the pairing is successful, the detection of the reference light within a flexible interval $T_r$ prior to the current moment is selected to estimate the frequency difference $\Delta f$ based on FFT. Based on this frequency difference, the keys can be distilled without reducing efficiency.}
    \label{fig_post-phase-locking}
\end{figure}

\begin{figure}
    \centering
    \includegraphics[width=0.95\textwidth]{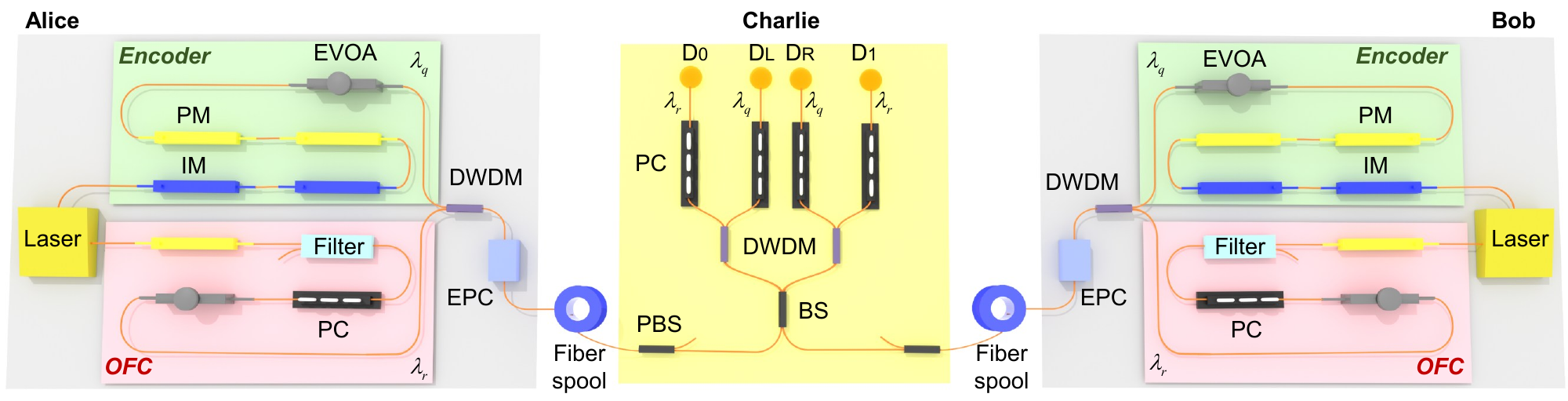}
    \caption{Experimental setup. Alice and Bob employ the free-running continuous-wave (CW) lasers without phase locking. The encoder unit is responsible for the preparation of encoding quantum states. It uses two intensity modulators (IMs) to chop the CW light into weak coherent pulses with three levels of intensity. These levels correspond to the signal state, the decoy state, and the vacuum state. Then the phase of the pulses is randomized with two phase modulators (PMs), and the intensity of the pulses is attenuated to the single-photon level with the electric variable optical attenuator (EVOA). The optical frequency comb (OFC) unit generates the reference light that is coherent with the quantum pulses but at a different wavelength. The electro-optic frequency comb is first generated by a PM. Then it is filtered with a narrowband filter to obtain a single sideband. Before coupling with the quantum pulses with a dense wavelength division multiplexer (DWDM), the polarization and intensity of the reference light are tuned to the proper levels with polarization controller (PC) and EVOA. At the output of the senders, the polarization drift in the fiber channel is pre-compensated with an electric polarization controller (EPC). Here, the intensity of the quantum pulses satisfies the settings specified by the protocol. Alice, Bob, and Charlie are connected with the standard single-mode fibers. Charlie is responsible for performing interference and detection on the quantum light and reference light, respectively. The polarization is first aligned with the polarization beam splitters (PBSs). The quantum light and reference light are interfered in a 50:50 beam splitter (BS), and are decoupled with the DWDMs. Last, the interference light is polarization-compensated to achieve high detection efficiency and is detected by four superconducting nanowire single-photon detectors (SNSPDs). Among them, $\text{D}_\text{L}$ and $\text{D}_\text{R}$ are used to detect the quantum light, while $\text{D}_0$ and $\text{D}_1$ are used to detect the reference light.}
    \label{fig_mp_experimental_setup}
\end{figure}

As illustrated in Fig. \ref{fig_mp_experimental_setup}, the experimental setup consists of two senders, Alice and Bob, and a measurement note, Charlie. Each sender employs a free-running continuous-wave (CW) laser with a central wavelength of 1549.72 nm (H34 channel) and a Lorentzian linewidth of 0.1 kHz (X15 laser, NKT Photonics Inc.). We use the sliding frequency referencing method to estimate and compensate for the frequency difference between Alice and Bob's lasers, which is shown in Fig. \ref{fig_post-phase-locking}.

The main fiber output of the laser is modulated by two cascaded intensity modulators (IMs) to generate a pulse train with three intensity levels, featuring a temporal width of 400 ps and a repetition rate of 500 MHz. The first IM chops the CW light into pulses of strong intensity or vacuum pulses. The second IM then modulates the strong intensity pulses into two levels to generate the signal or decoy pulses, while also enhancing the extinction ratio of the vacuum pulses. This setup ensures a dynamic extinction ratio of over 50 dB. Two phase modulators (PMs) are employed to randomize the phase of the pulses. Then the pulses are attenuated to the single-photon level using an electric variable optical attenuator (EVOA). The quantum pulses are coupled with the reference light using a DWDM. Before transmitting to Charlie, the polarization is pre-compensated to compensate for the polarization drift in the quantum channel, utilizing an electric polarization controller (EPC) based on the feedback from Charlie. At the output of the EPC, the quantum pulse intensities meet the requirements specified by the MP-QKD protocol.

The monitor optical output of the laser is modulated by a PM driven by a 25 GHz microwave signal to generate an electro-optic frequency comb. The electro-optic frequency comb can be shown as \cite{zhuang2023RN1102}
\begin{equation}
    \begin{aligned}
        E_{\text{comb}} = E_0 \Big\{ & J_0 (\pi k_m) \cos (2\pi f_0 t + \varphi_0) + (-1)^n \sum_{n=1}^\infty J_n (\pi k_m) \cos[2\pi (f_0 + n f_m)t + n \varphi_m + \varphi_0]\\
        & + \sum_{n=1}^\infty J_n (\pi k_m) \cos[2\pi (f_0 - n f_m)t + n \varphi_m + \varphi_0] \Big\},
    \end{aligned}
\end{equation}
where $E_0$ is the amplitude of input light, $J_n$ is the Bessel function of $n$ order, $k_m = V_m/V_\pi$ with the amplitude of modulation signal $V_m$ and the half-wave voltage of the PM $V_\pi$, $f_0$ is the frequency of the input light, $t$ is the time, $\varphi_0$ is the initial phase of the input light, $f_m$ is the frequency of the modulation signal, and $\varphi_m$ is the initial phase of the modulation signal. The amplitude of the sidebands obeys the Bessel function of different orders and the spacing between adjacent sidebands is equal to the frequency of the modulation signal $f_m$. As the frequency of the modulation signal is set as 25 GHz, the spacing between the 0th-order and 2nd-order sidebands is 50 GHz. We use the filter to retain a single second-order sideband (1550.12nm, C34 channel) as a frequency reference. We use this frequency reference light to establish coherence between Alice and Bob's lasers. The frequency difference of the second-order sidebands between Alice and Bob is just the frequency difference of the original lasers. This guarantees the efficiency of the simple sliding frequency referencing method. The frequency reference light undergoes polarization compensation using a polarization controller (PC) and is then attenuated to an appropriate intensity level using a variable optical attenuator (VOA). Finally, it is coupled with the quantum light pulses via a DWDM.

The channel is the standard single-mode fiber with a typical loss coefficient of 0.2 dB/km. At the receiver, the quantum light pulses and the CW frequency reference light first pass through the polarization beam splitters (PBSs) to ensure identical polarization. Then they interfere at the same 50:50 beam splitter (BS). The quantum light pulses and the CW frequency reference light are decoupled using the DWDMs. After polarization compensation with PCs, they are detected using the superconducting nanowire single-photon detectors (SNSPDs). The detection efficiency and dark count rate of the SNSPDs are calibrated to approximately 55\% and 30 Hz. Based on the detection results of the frequency reference light, we estimate the beat note $\Delta f$ based on FFT. Then Alice and Bob calculate the relative phase difference in paired bins $j$ and $k$ and sift the pairs in X basis satisfying
\begin{equation}
    \Big|\delta_{jk}^a - \delta_{jk}^b + \frac{2 \pi}{F} \Delta f (k - j) \Big| \leq \frac{2\pi}{M} \ \text{or} \ \Big|\Big|\delta_{jk}^a - \delta_{jk}^b + \frac{2 \pi}{F} \Delta f (k - j)\Big| - \pi\Big| \leq \frac{2\pi}{M},
\end{equation}
where $F$ is the clock rate of the MP-QKD system. Then Alice and Bob process the detection results of the quantum light pulses to extract secret keys.

The stable operation of the whole system is achieved through coordinated tracking and compensation among Alice, Bob, and Charlie. Synchronization is achieved by electrically distributing a 10 MHz clock, generated by a field-programmable gate array (FPGA) at Charlie, to Alice and Bob. To compensate for the polarization drift in fiber links, Charlie provides feedback on the counting rate of the quantum light pulses via Ethernet every 300 ms. Alice and Bob alternately adjust their polarization with EPCs to maximize the counting rate. We can perform this compensation because the wavelength difference between Alice's and Bob's lasers is on the order of MHz. In addition, Charlie calculates the delay between Alice and Bob by analyzing the count distribution and provides feedback to Alice via Ethernet for compensation. More details are present in the App. \ref{system_design}.

\section{Experimental results}

\begin{figure}
    \centering
    \includegraphics[width=0.9\textwidth]{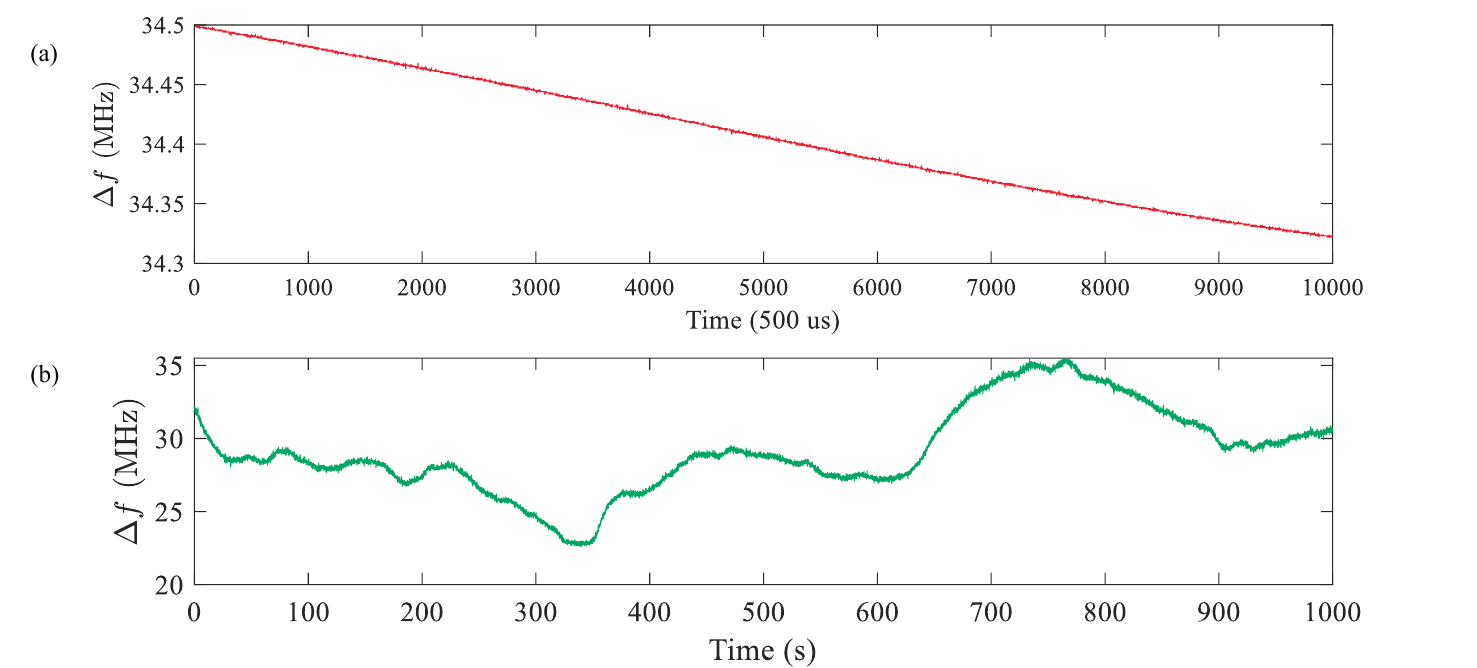}
    \caption{The beat frequency between two free-running lasers without fiber spool. (a) The short-period beat frequency is measured using SNSPDs every 500 us. (b) The long-period beat frequency is measured using classical photodetectors every 1 second.}
    \label{fig_fft_result}
\end{figure}

We first pre-calibrate the two remote lasers. The central wavelengths are set as 1549.72 nm and the beat frequency is measured as around 34 MHz. In Fig. \ref{fig_fft_result}(a), the beat frequency is measured using SNSPDs every 500 us by padding zero to double the length, which clearly exhibits a frequency drift in 5 seconds. The resolution of the FFT is 1000 Hz. To characterize the relative stability of the two lasers, we calculate the difference between consecutive FFT results and obtain the standard deviation as 652.282 Hz. The stability on the microsecond timescale is enough to perform MP-QKD. If we consider a maximum pairing interval ($L_{\max} = 50000$, i.e., 100 us) in our experiments, the maximum residual phase due to laser instability and FFT precision issues is $\delta_{\text{LD}} = 0.3961$. The error rate in X basis with residual phase $\delta$ is given as
\begin{equation}
    E_X (\delta) = \frac{1}{2\pi} \int_0^{2\pi} \cos^2 \bigg(\frac{\theta}{2}\bigg) \sin^2 \bigg(\frac{\theta + \delta}{2}\bigg) + \sin^2 \bigg(\frac{\theta}{2}\bigg) \cos^2 \bigg(\frac{\theta+\delta}{2}\bigg) d\theta = \frac{1}{4} (2 - \cos\delta).
\end{equation}
When there is no residual phase, i.e., $\delta = 0$, the error rate in X basis is 25\%. If we consider the maximum residual phase $\delta_{\text{LD}}$, the error rate is $E_X (\delta_{\text{LD}}) = 26.94\%$. We note that this is the worst case, the average pairing interval is shorter and the average accuracy is higher. Therefore, we could perform the MP-QKD experiments based on this setup. In Fig. \ref{fig_fft_result}(b), we measure the beat frequency with a strong light, which reveals a noticeable frequency drift over time. The detailed measurement methods are given in the App. \ref{frequency_stability}. We note that under these conditions, it is not feasible to directly conduct MP-QKD experiments. However, the long-period frequency drift will not affect our sliding frequency referencing method, as it only requires stability on the microsecond timescale. Furthermore, our method necessitates only a one-time calibration of the beat frequency at the start. This initial calibration ensures stable operation without significant deviations. But no further calibration is required during QKD.

Then we measure the relative phase drift over long channels. Although MP-QKD is inherently robust against phase drift, excessive phase drift can still introduce unnecessary errors. The standard deviations of the relative phase drift are 0.0987, 0.0973, 0.1155, and 0.1489 rad per 100 us over 202.31, 303.37, 354.62, and 404.25 km, respectively. The details are present in the App. \ref{phase_drift}. Considering a maximum pairing interval ($L_{\max} = 50000$, i.e., 100 us), the residual phase caused by phase drift is $\delta_{\text{pd}}=$ 0.1489 rad. On average, the X basis error rate is only a little higher as $E_X (\delta_{\text{pd}}) = $25.28\%. It should be noted that the residual phase error discussed here corresponds to the maximum pairing length. In practice, the average pairing interval is shorter, resulting in a relative phase drift that is small enough to induce negligible errors.

We implement the experiment with the standard single-mode fiber with a typical loss coefficient of 0.2 dB/km. We optimize the probabilities $p_{\mu_a}, p_{\nu_a}$ and the intensities $\mu_a, \nu_a$ at every distance by limiting that $p_{\mu_a} + p_{\nu_a} < 1$, $\mu_a > \nu_a$, and $0.2 > \nu_a > 0.01$. As we consider the symmetric link condition, Bob's parameters are set the same as Alice's. To obtain accurate frequency reference, we set the interval $T_r = 500$ us and pad additional zeros to double its length, which guarantees a resolution of 1000 Hz. The average counting rate of the reference light is greater than 1 MHz at all distances. We optimize the maximum pairing interval $L_{\max}$ from 10000 to 50000 to balance the pairing efficiency and the error rate in X basis. Besides, the block length $N$ are accumulated from $1.38 \times 10^{12}$ to $6.27 \times 10^{13}$ to mitigate the finite key effects. The detailed settings are listed in the App. \ref{detailed_experimental_results}.

\begin{figure}[t]
    \centering
    \includegraphics[width=0.64\textwidth]{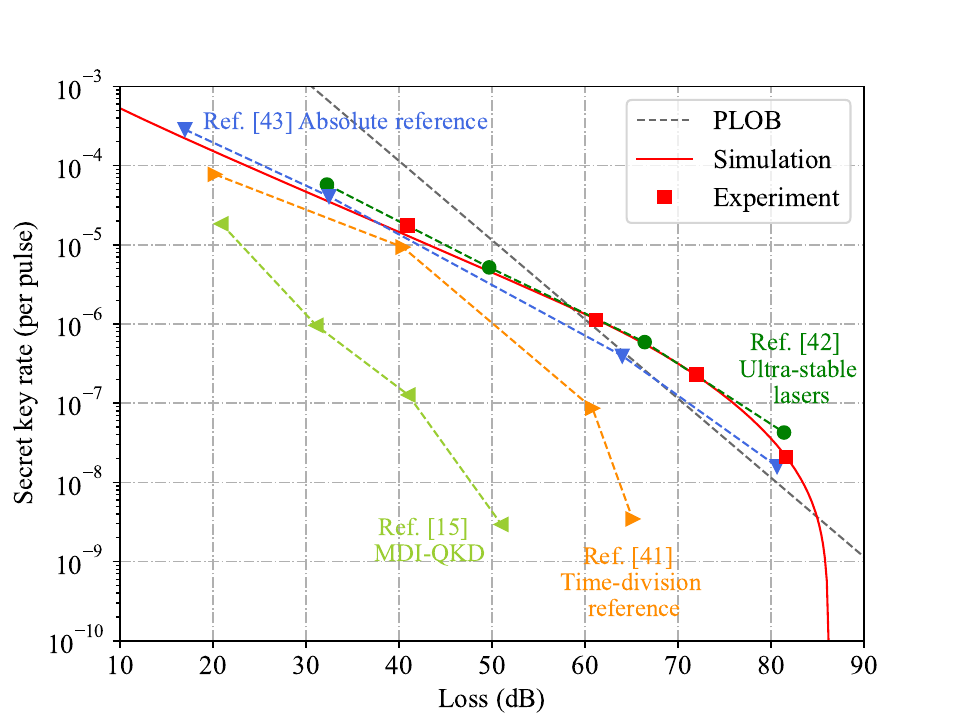}
    \caption{Secret key rate in logarithmic scale versus transmission loss between Alice and Bob. The red squares represent the experimental results obtained at distances of 202.31, 303.37, 354.62, and 404.25 km, corresponding to transmission losses of 40.92, 61.19, 72.01, and 81.60 dB, respectively. The red line represents the simulation results with settings at 404.25 km. The fiber loss coefficient is set typically as 0.2 dB/km. The PLOB bound \cite{pirandola2017RN103} and the recent QKD experiments \cite{yin2016RN288,zhu2023RN928,zhou2023RN978,ge2025RN1181} are shown for comparison.}
    \label{fig_mp_skr}
\end{figure}

The experimental SKR results are shown in Fig. \ref{fig_mp_skr} at the transmission distances of 202.31, 303.37, 354.62, and 404.25 km. The X-axis in Fig. \ref{fig_mp_skr} is plotted in terms of transmission loss. The transmission losses are 40.92, 61.19, 72.01, and 81.60 dB, respectively. The simulated SKR based on the settings at 404.25 km and the rate-loss limit (i.e., PLOB bound \cite{pirandola2017RN103}) are also shown in Fig. \ref{fig_mp_skr}. The results show that the obtained SKRs surpass the rate-loss limit at 303.37, 354.62, and 404.25 km, which are 1.0267, 2.5230, and 2.1033 times of the PLOB bound. 

We compare the obtained SKRs with some recent QKD experiments. Compared with traditional MDI-QKD \cite{yin2016RN288}, MP-QKD shows significant advancements in terms of both SKR and transmission distance. Our frequency-comb-based scheme exhibits superior performance compared to the time-division reference approach \cite{zhu2023RN928}. Besides, our results outperform those of the local absolute frequency method at long distance \cite{ge2025RN1181}. Even in comparison to the MP-QKD experiments based on ultra-stable lasers \cite{zhou2023RN978}, our method demonstrates performance at an almost equivalent level. At distances of 202.31, 303.37, and 354.62 km, our experimental results align closely with the trend line in Ref. \cite{zhou2023RN978}. However, due to the lower stability of our lasers compared to the ultra-stable lasers, which limit the maximum pairing interval, our result falls slightly below that of Ref. \cite{zhou2023RN978} at 404.25 km. These results validate that our method is not only convenient in implementation but also highly efficient in performance, making it suitable for practical applications.

\section{Conclusion}

In this work, we demonstrate a simplified yet high-performance MP-QKD system based on optical frequency combs, eliminating the need for active frequency locking technology. By generating and utilizing different sidebands from the same comb sources, we design a sliding frequency referencing method to achieve both coherence establishment and key distribution within a single fiber channel. Our system significantly simplifies the implementation complexity and demonstrates the capability of MP-QKD to surpass the rate-loss limit using free-running commercial lasers. In future work, we expect to improve the clock rate, which could yield a more-than-proportional improvement in SKR. Overall, this study represents a significant step forward in simplifying high-performance QKD systems and paves the way for practical and scalable long-distance communication.

\clearpage

\appendix

\section{Pairing Strategy}

We present the pairing strategy that first filters out ineffective rounds and subsequently pairs adjacent effective rounds within the maximum possible pairing interval \cite{zhou2023RN978,xie2023RN1097,lu2025RN1293}.

\begin{algorithm}[H]
    \SetAlgoLined
    \KwIn{Alice and Bob's intensities $\tau_j^a$ and $\tau_j^b$, and Charlie's announced detection results $C_j = L_j \oplus R_j$ for $j=1$ to $N$; maximum pairing interval $L_{\max}$.}
    \KwOut{$D$ pairs, $(F_d, R_d)$, as the $d$-th pair for $d=1$ to $D$.}
    Initialize the index $d=1$, the flag $f=0$; Initialize $C_j^\prime = 0$ for $j=1$ to $N$\;
    \For{$j\leftarrow 1$ \KwTo $N$}{
        \eIf{$(\tau_j^a,\tau_j^b) = (\mu_j^a,\nu_j^b)$ or $(\nu_j^a,\mu_j^b)$}{
            $C_j^\prime \leftarrow 0$\;
        }{
            $C_j^\prime \leftarrow C_j$\;
        }
    }
    \For{$j\leftarrow 1$ \KwTo $N$}{
        \If{$C_j^\prime = 1$}{
            \eIf{$f = 0$}{
                $F_d \leftarrow j$; $f \leftarrow 1$\;
            }{
                \eIf{$j - F_k \leq L_{\max}$}{
                    $R_d \leftarrow j$; $d \leftarrow d + 1$; $f \leftarrow 0$\;
                }{
                    $F_d \leftarrow j$;
                }
            }
        }
    }
    Set the number of pairs as $D \leftarrow d - 1$.
    \caption{Pairing Strategy}
    \label{pairing_strategy}
\end{algorithm}

\section{Parameter estimation method}

We employ the decoy-state method \cite{xu2013RN455,zhou2023RN978} to estimate the bounds of the single-photon states, i.e., $\underline{n}_{11}^z$ and $\overline{e}_{11}^{\text{ph}}$. First, we introduce some basic definitions. Denote the number of pairs as $n_{[\tau_a,\tau_b]}$, where $\tau_a = \tau_j^a + \tau_k^a$ and $\tau_b = \tau_j^b + \tau_k^b$ in a pair. And denote the number of error pairs in X basis as $m_{[2\nu_a,2\nu_b]}$. For those observed values denoted as $n$, we could estimate the lower and upper bound of the expected value with success probabilities $1-\varepsilon_U$ and $1-\varepsilon_L$ as
\begin{equation}
    \begin{aligned}
        \overline{n} &= \frac{n}{1 + \chi_U},\\
        \underline{n} &= \frac{n}{1 + \chi_L},
    \end{aligned}
\end{equation}
where $\chi_U$ and $\chi_L$ are the solution of the following equations
\begin{equation}
    \begin{aligned}
        \bigg[ \frac{e^{-\chi_U}}{(1 - \chi_U)^{1 - \chi_U}} \bigg]^{\frac{n}{1 - \chi_U}} = \varepsilon_U,\\
        \bigg[ \frac{e^{\chi_L}}{(1 + \chi_L)^{1 + \chi_L}} \bigg]^{\frac{n}{1 + \chi_L}} = \varepsilon_L.
    \end{aligned}
\end{equation}
Define $p_{[\tau_a,\tau_b]}$ as the probability of the pairs corresponding to $[\tau_a,\tau_b]$ among all pairs, which can be expressed as
\begin{equation}
    p_{[\tau_a,\tau_b]} = \frac{1}{p_s} \sum_{\tau_j^a + \tau_k^a = \tau_a} \sum_{\tau_j^b + \tau_k^b = \tau_b} p_{\tau_j^a} p_{\tau_k^a} p_{\tau_j^b} p_{\tau_k^b},
\end{equation}
where $p_s$ is the total probability and will not affect the subsequent calculations. A particular case is the pairs $[2\nu_a,2\nu_b]$, which is additionally sifted according to the relative phase difference, and can be shown as
\begin{equation}
    p_{[2\nu_a,2\nu_b]} = \frac{2}{M} \frac{1}{p_s} p_{\nu_j^a}^2 p_{\nu_j^b}^2.
\end{equation}

Define two linear combinations of the number of pairs as follows
\begin{equation}
    \begin{aligned}
        n_\mu &\triangleq \frac{\underline{n}_{[\mu_a,\mu_b]}}{p_{0,\mu_a} p_{0,\mu_b} p_{[\mu_a,\mu_b]}} - \frac{\overline{n}_{[o_a,\mu_b]}}{p_{0,o_a} p_{0,\mu_b} p_{[o_a,\mu_b]}} - \frac{\overline{n}_{[\mu_a,o_b]}}{p_{0,\mu_a} p_{0,o_b} p_{[\mu_a,o_b]}} + \frac{\underline{n}_{[o_a,o_b]}}{p_{0,o_a} p_{0,o_b} p_{[o_a,o_b]}},\\
        n_\nu &\triangleq \frac{\overline{n}_{[\nu_a,\nu_b]}}{p_{0,\nu_a} p_{0,\nu_b} p_{[\nu_a,\nu_b]}} - \frac{\underline{n}_{[o_a,\nu_b]}}{p_{0,o_a} p_{0,\nu_b} p_{[o_a,\nu_b]}} - \frac{\underline{n}_{[\nu_a,o_b]}}{p_{0,\nu_a} p_{0,o_b} p_{[\nu_a,o_b]}} + \frac{\underline{n}_{[o_a,o_b]}}{p_{0,o_a} p_{0,o_b} p_{[o_a,o_b]}},
    \end{aligned}
\end{equation}
where $p_{i,\tau} = e^{-\tau} \tau^i / i!$ is the Poisson distribution probability. Then the lower bound of the number of the single-photon pairs in Z basis can be estimated as
\begin{equation}
    \underline{n}_{11}^z = \frac{1}{\alpha_{11}} \bigg[ \frac{p_{0,\nu_a} p_{0,\nu_b}}{p_{s_a,\nu_a} p_{s_b,\nu_b}} n_\nu - \frac{p_{0,\mu_a} p_{0,\mu_b}}{p_{s_a,\mu_a} p_{s_b,\mu_b}} n_\mu \bigg],
\end{equation}
where $s_a=1, s_b=2$ when $\nu_a \mu_b \leq \nu_b \mu_a$, $s_a=2, s_b=1$ when $\nu_a \mu_b > \nu_b \mu_a$, and $\alpha_{11}$ is defined as
\begin{equation}
    \alpha_{11} = \frac{1}{p_{[\mu_a,\mu_b]}} \bigg[\frac{p_{1,\nu_a} p_{1,\nu_b}}{p_{s_a,\nu_a} p_{s_b,\nu_b} p_{1,\mu_a} p_{1,\nu_b}} - \frac{1}{p_{s_a,\mu_a} p_{s_b,\mu_b}} \bigg].
\end{equation}

To estimate the phase error rate, we define a linear combination as
\begin{equation}
    m_{2\nu} \triangleq \frac{m_{[2\nu_a,2\nu_b]}}{p_{0,2\nu_a} p_{0,2\nu_b} p_{[2\nu_a,2\nu_b]}} - \frac{\underline{n}_{[o_a,2\nu_b]}}{2p_{0,o_a} p_{0,2\nu_b} p_{[o_a,2\nu_b]}} - \frac{\underline{n}_{[2\nu_a,o_b]}}{2p_{0,2\nu_a} p_{0,o_b} p_{[2\nu_a,o_b]}} + \frac{\overline{n}_{[o_a,o_b]}}{2p_{0,o_a} p_{0,o_b} p_{[o_a,o_b]}},
\end{equation}
and the upper bound of the error single-photon pairs in X basis can be estimated as
\begin{equation}
    \overline{m}_{11,[2\nu_a,2\nu_b]} = p_{0,2\nu_a} p_{0,2\nu_b} p_{[2\nu_a,2\nu_b]} m_{2\nu}.
\end{equation}
The lower bound of the total number of single-photon pairs in X basis is
\begin{equation}
    \underline{n}_{11,[2\nu_a,2\nu_b]} = \frac{p_{1,2\nu_a} p_{1,2\nu_b} p_{[2\nu_a,2\nu_b]}}{p_{1,\mu_a} p_{1,\mu_b} p_{[\mu_a,\mu_b]}} \underline{n}_{11}^z.
\end{equation}
Therefore, the upper bound of the single-photon error rate in X basis is calculated as 
\begin{equation}
    \overline{e}_{11}^x = \frac{\overline{m}_{11,[2\nu_a,2\nu_b]}}{\underline{n}_{11,[2\nu_a,2\nu_b]}}.
\end{equation}
Based on the random-sampling theory without replacement, the upper bound of the phase error rate is estimated with probability at least $1 - \varepsilon_{e}$ \cite{lim2014RN93,chau2018RN77}
\begin{equation}
    \overline{e}_{11}^{\text{ph}} = \overline{e}_{11}^x + \gamma \big(\varepsilon_{e}, \overline{e}_{11}^x, \underline{n}_{11}^z, \underline{n}_{11,[2\nu_a,2\nu_b]} \big),
\end{equation}
where
\begin{equation}
    \gamma(a,b,c,d) = \sqrt{\frac{(c+d)(1-b)b}{cd} \ln \Big[ \frac{c+d}{2\pi cd (1-b) b a^2} \Big]}.
\end{equation}

\section{System Design}
\label{system_design}

\begin{figure}
    \centering
    \includegraphics[width=0.9\textwidth]{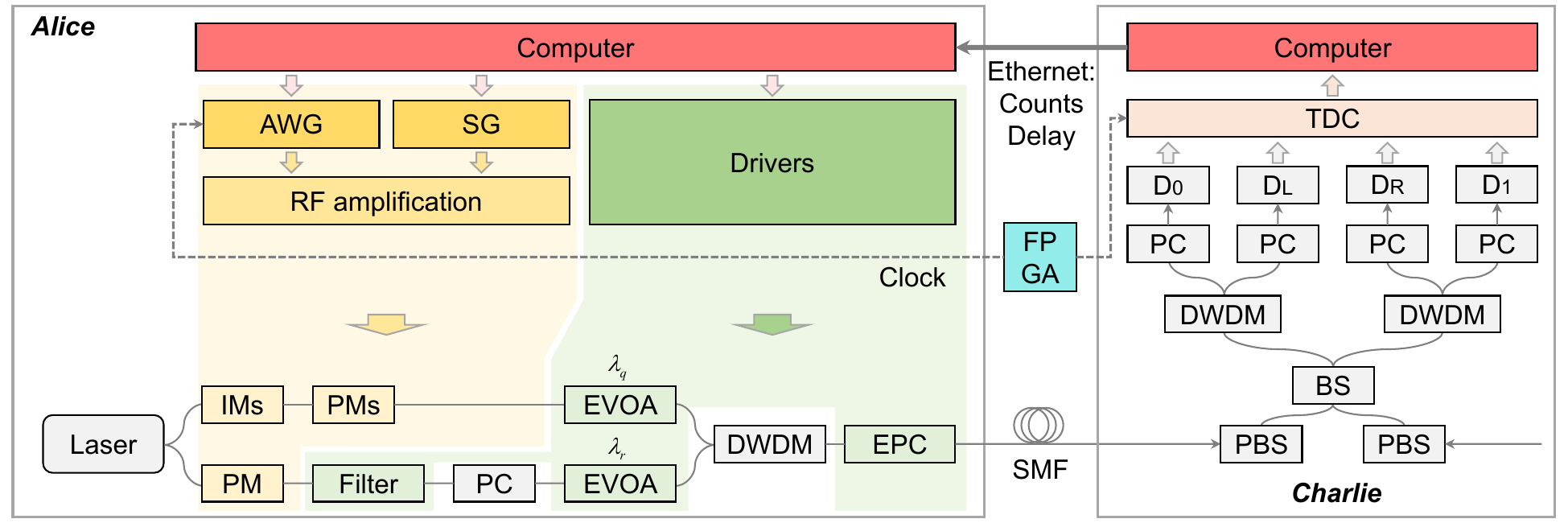}
    \caption{Schematic diagram of the MP-QKD experimental setup at Alice and Charlie's sides. Bob's setup is the same as Alice's. The entire QKD system is centrally controlled by the computers. The devices marked in orange or cyan are driven and controlled by the devices of the same color. The devices labeled in grey are not subject to control or manual tuning. AWG, arbitrary waveform generator; SG, signal generator; IM, intensity modulator; PM, phase modulator; PC, polarization controller; EVOA, electric variable optical attenuator; DWDM, dense wavelength division multiplexer; EPC, electric polarization controller; SMF, single-mode fiber; FPGA, field programmable gate array; PBS, polarization beam splitter; BS, beam splitter; TDC, time-to-digital converter; $\text{D}_0$, $\text{D}_1$, $\text{D}_\text{L}$, and $\text{D}_\text{R}$ are the detectors.}
    \label{fig_system_design}
\end{figure}

In Fig. \ref{fig_system_design}, we show the schematic diagram of the whole experimental setup. The modulation signals on the quantum light $\lambda_q$ are generated by a 25 GSa/s arbitrary waveform generator (AWG) and amplified by 20 GHz RF amplifiers. The 25 GHz microwave signal on the PM to generate the optical frequency comb is generated by a 44 GHz signal generator and amplified by an RF amplifier. The filter, electric variable optical attenuators (EVOAs), and electric polarization controller (EPC) are driven by the dedicated drivers. The three parties are synchronized by referencing the 10 MHz clock generated by a field programmable gate array (FPGA). This electrical signal synchronization method can be upgraded to optical synchronization for practical applications. To compensate for the slow polarization drift in the fiber channels, the pre-compensation is performed at the transmitter using an electric polarization controller (EPC). The reference for polarization compensation is the real-time count of $\text{D}_\text{L}$ per 300 ms from Charlie via the Ethernet. Here, we use the polarization beam splitter (PBS) to align the polarization and convert the polarization information into intensity information for polarization compensation. Besides, the light from Alice and Bob will not interfere as the frequency difference is at the MHz level. Therefore, we could directly use this count of $\text{D}_\text{L}$ as the reference for polarization compensation. Since polarization drift is slow, Alice and Bob can alternately perform polarization compensation in the same way. This setup helps to mitigate the loss at the receiver. To compensate for the slow time delay between Alice and Bob's quantum pulses, Charlie evaluates the timing delay difference between Alice and Bob based on the statistical information from detector $\text{D}_\text{L}$ and feeds back it only to Alice via the Ethernet every 1000 s. Then Alice aligns the delay of the modulation signals directly on the AWG.

\section{Beat frequency characterisation}
\label{frequency_stability}

\begin{figure}
    \centering
    \includegraphics[width=0.30\textwidth]{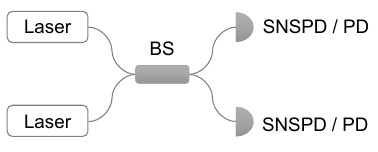}
    \caption{Beat frequency measurement scheme. The attenuators are omitted. BS, beam splitter; SNSPD, superconducting nanowire single-photon detector; PD, photodetector.}
    \label{fig_beat}
\end{figure}

\begin{figure}
    \centering
    \includegraphics[width=1\textwidth]{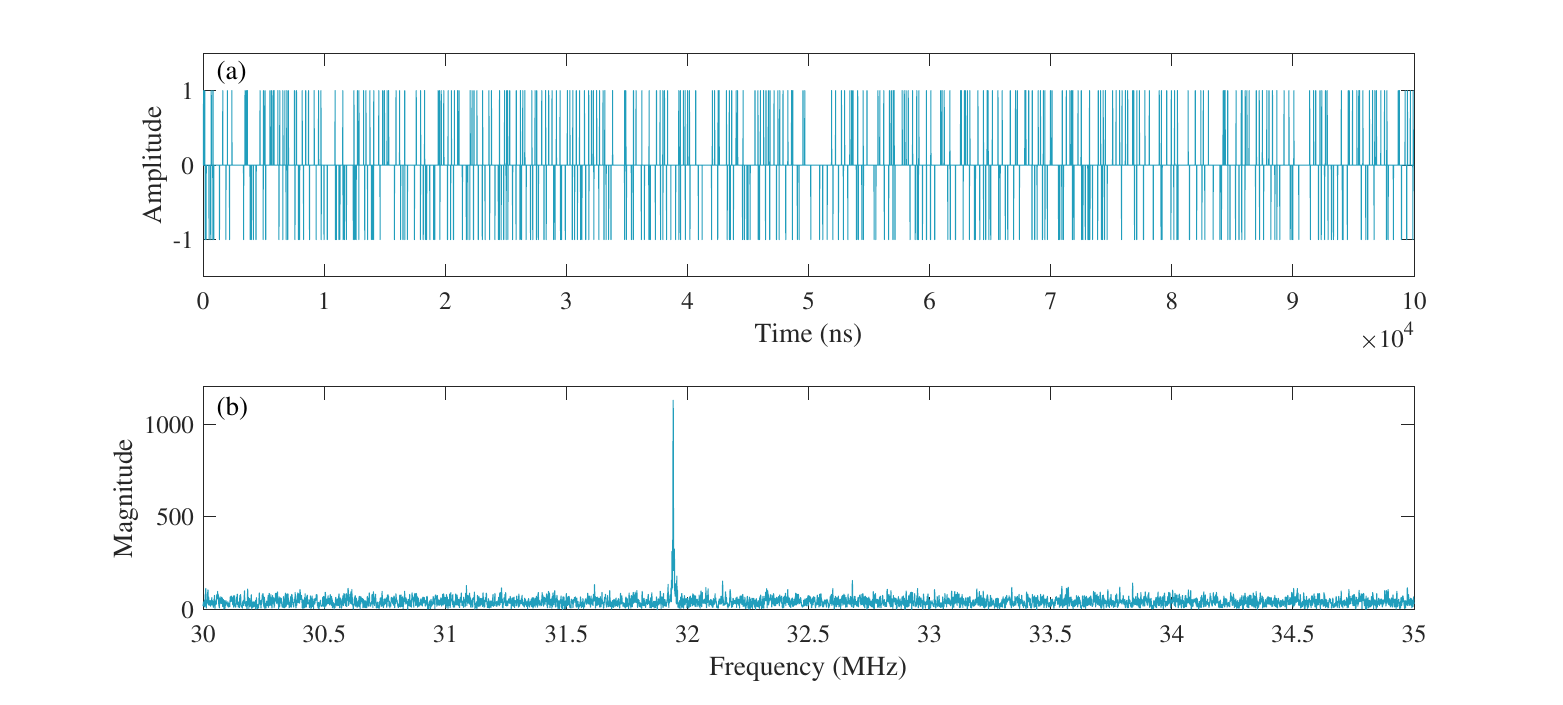}
    \caption{The results of FFT analysis. (a) The partial time-domain signal. (b) The partial frequency spectrum obtained after FFT.}
    \label{fig_fft_in_out}
\end{figure}

We measured the beat frequency under both strong and weak light conditions using the experimental scheme illustrated in Fig. \ref{fig_beat}. Under strong light conditions, we use one PD to convert the beat frequency optical signal into an electrical signal for direct frequency analysis. Under the weak light condition, we first attenuate the light to a level suitable for SNSPDs and then measure the beat frequency results with two channels of SNSPD. We use FFT to calculate the beat frequency every 500 us. Fig. \ref{fig_fft_in_out} shows an example of the FFT analysis. The time-domain signals are obtained with a resolution of 1 ns in Fig. \ref{fig_fft_in_out}(a). The magnitude distribution as a function of frequencies is shown in Fig. \ref{fig_fft_in_out}(b). Here, Fig. \ref{fig_fft_in_out} presents only part of the FFT results for better visualization.

\section{Phase drift characterisation}
\label{phase_drift}

\begin{figure}
    \centering
    \includegraphics[width=0.46\textwidth]{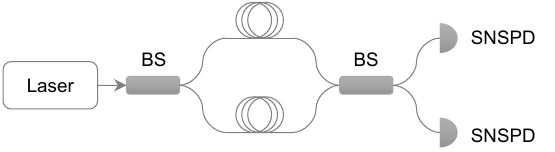}
    \caption{Phase drift measurement scheme. The attenuators are omitted. BS, beam splitter; SNSPD, superconducting nanowire single-photon detector.}
    \label{fig_phase_drift_measurement}
\end{figure}

\begin{figure}[t]
    \centering
    \includegraphics[width=0.96\textwidth]{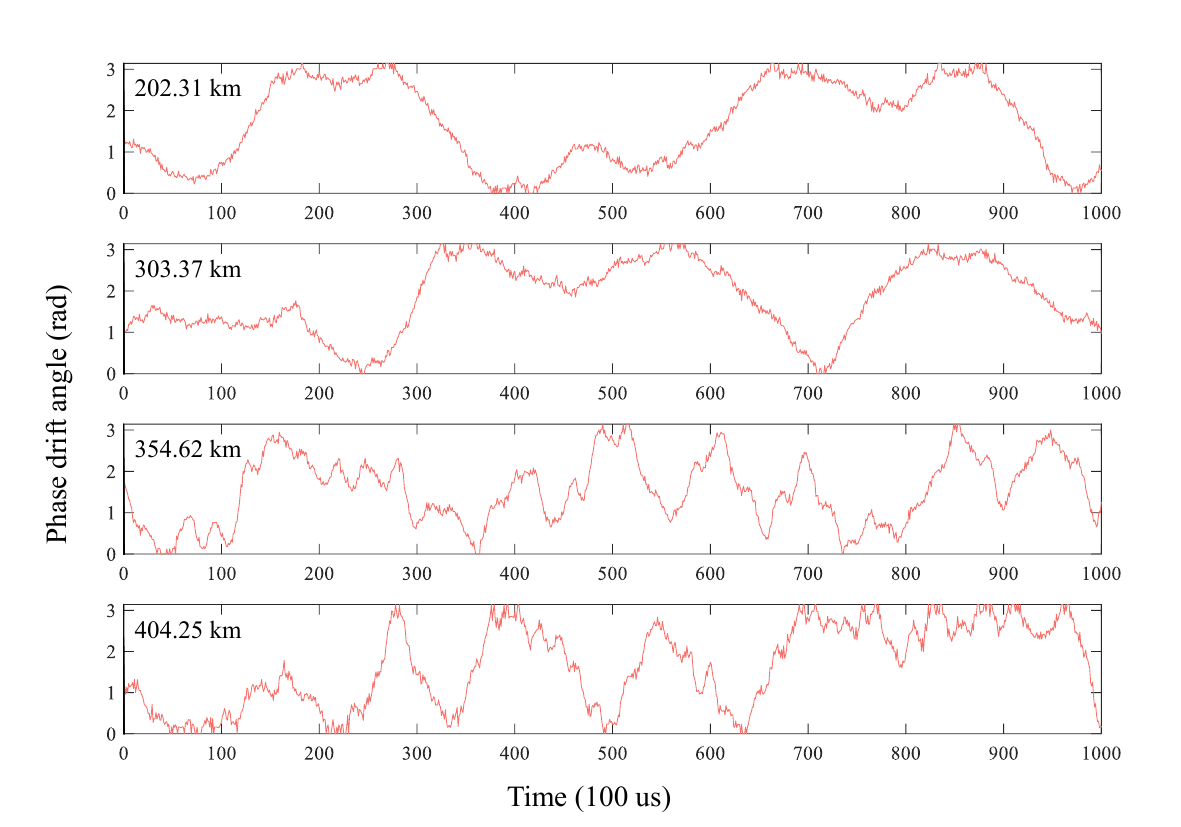}
    \caption{The phase drift measured every 100 us at distances of 202.31, 303.37, 354.62, and 404.25 km.}
    \label{fig_phase_drift}
\end{figure}

\begin{figure}[t]
    \centering
    \includegraphics[width=0.96\textwidth]{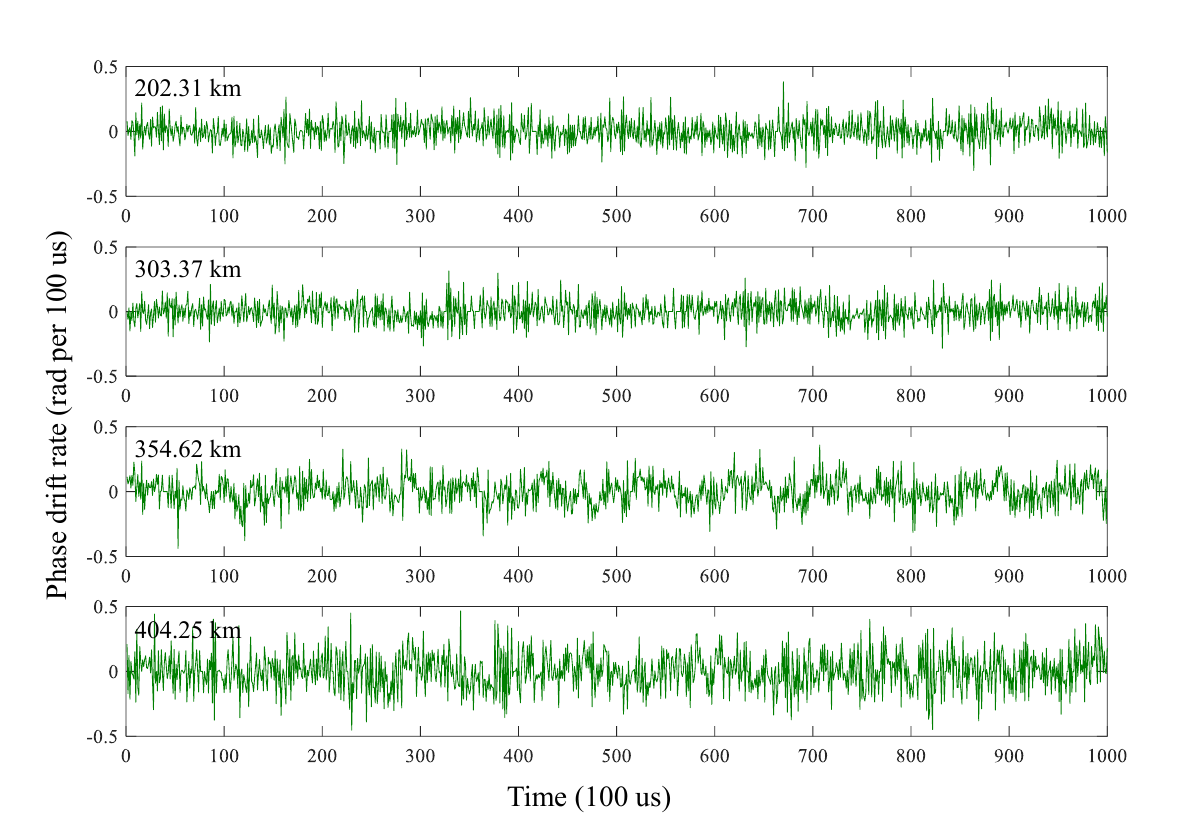}
    \caption{The relative phase drift measured every 100 us at distances of 202.31, 303.37, 354.62, and 404.25 km. The standard deviations of the relative phase drift are 0.0987, 0.0973, 0.1155, and 0.1489 rad per 100 us, respectively.}
    \label{fig_relative_phase_drift}
\end{figure}

We measured the phase drift with the experimental scheme illustrated in Fig. \ref{fig_phase_drift_measurement}. Denote the counts per 100 us of two SNSPDs as $C_0$ and $C_1$. We adjust the attenuators to guarantee the counts of SNSPDs satisfy that $C_0 + C_1 \geq 600$. Then the phase drift angle $\theta_{\text{pd}}$ at interval $[0,\pi]$ can be calculated as
\begin{equation}
    \theta_{\text{pd}} = \arccos \bigg( \frac{C_0 - C_1}{C_0 + C_1} \bigg).
\end{equation}
The results of the phase drift at 202.31, 303.37, 354.62, and 404.25 km are shown in Fig. \ref{fig_phase_drift}. The relative phase drift rates are shown in Fig. \ref{fig_relative_phase_drift}. The standard deviations of the relative phase drift are 0.0987, 0.0973, 0.1155, 0.1489 rad per 100 us.

\section{Detailed experimental parameters and results}
\label{detailed_experimental_results}

We present the experimental parameters and results of MP-QKD in Tabs. \ref{tab_exp_parameters} and \ref{tab_exp_results}. In Tab. \ref{tab_exp_parameters}, the intensities $\mu_a,\mu_b,\nu_a,\nu_b$ are experimentally measured values, and therefore Alice and Bob's settings differ slightly. Additionally, because pseudo-random numbers are employed in the experiment, Alice and Bob's statistical probabilities $p_{\mu_a}, p_{\nu_a}, p_{o_a}, p_{\mu_b}, p_{\nu_b}, p_{o_b}$ also exhibit slight differences.

\setlength{\tabcolsep}{8mm}
\begin{table}[t]
    \centering
    \caption{Experimental parameters of MP-QKD at different fiber lengths. $L_{AC}$ ($L_{BC}$) is the fiber length between Alice (Bob) and Charlie. $\alpha_{AC}$ ($\alpha_{BC}$) is the transmission loss between Alice (Bob) and Charlie. The total loss is the sum of $\alpha_{AC}$ and $\alpha_{BC}$.}
    \begin{tabular}{ccccc}
    \toprule
    Distance   (km)     & 202.31          & 303.37          & 354.62          & 404.25          \\ \midrule
    $L_{AC}$ (km)       & 101.14          & 151.68          & 177.17          & 202.12          \\
    $L_{BC}$ (km)       & 101.17          & 151.69          & 177.45          & 202.14          \\ \midrule
    Total Loss (dB)     & 40.92           & 61.19           & 72.01           & 81.60           \\
    $\alpha_{AC}$ (dB)  & 20.65           & 31.19           & 36.80           & 41.23           \\
    $\alpha_{BC}$ (dB)  & 20.27           & 30.00           & 35.21           & 40.37           \\ \midrule
    $\mu_a$             & 0.5216          & 0.4599          & 0.4028          & 0.4486          \\
    $\mu_b$             & 0.5256          & 0.4569          & 0.3910          & 0.4490          \\
    $\nu_a$             & 0.0487          & 0.0452          & 0.0307          & 0.0335          \\
    $\nu_b$             & 0.0489          & 0.0451          & 0.0301          & 0.0332          \\ \midrule
    $p_{\mu_a}$         & 0.2844          & 0.2458          & 0.2518          & 0.2518          \\
    $p_{\mu_b}$         & 0.2812          & 0.2450          & 0.2584          & 0.2584          \\
    $p_{\nu_a}$         & 0.2572          & 0.2167          & 0.3062          & 0.3062          \\
    $p_{\nu_b}$         & 0.2551          & 0.2078          & 0.3109          & 0.3109          \\
    $p_{o_a}$           & 0.4584          & 0.5375          & 0.4420          & 0.4420          \\
    $p_{o_b}$           & 0.4637          & 0.5472          & 0.4307          & 0.4307          \\ \midrule
    Clock (Hz)          & $5 \times 10^8$ & $5 \times 10^8$ & $5 \times 10^8$ & $5 \times 10^8$ \\ \bottomrule
    \end{tabular}
    \label{tab_exp_parameters}
\end{table}

\setlength{\tabcolsep}{4mm}
\begin{table}[t]
    \centering
    \caption{Experimental results of MP-QKD at different fiber lengths. $E_{[\mu_a,\mu_b]}$ and $E_{[2\nu_a,2\nu_b]}$ are the error rate of the pairs $[\mu_a,\mu_b]$ and $[2\nu_a,2\nu_b]$, respectively. The SKRs are shown in units of bits per pulse (bpp) and bits per second (bps), respectively.}
    \begin{tabular}{ccccc}
    \toprule
    Distance   (km)                 & 202.31                  & 303.37                  & 354.62                  & 404.25                  \\ \midrule
    $N$                             & $1.38 \times 10^{12}$   & $1.07 \times 10^{13}$   & $2.40 \times 10^{13}$   & $6.27 \times 10^{13}$   \\
    $L_{\max}$                      & 10000                   & 20000                   & 40000                   & 50000                   \\ 
    $T_r$ (us)                      & 500                     & 500                     & 500                     & 500                     \\ \midrule
    $n_{[\mu_a,\mu_b]}$             & 115886048               & 80563994                & 31730520                & 16433010                \\
    $m_{[\mu_a,\mu_b]}$             & 82970                   & 173723                  & 249295                  & 210473                  \\
    $n_{[\mu_a,o_b]}$               & 67139                   & 204522                  & 218132                  & 177056                  \\
    $n_{[o_a,\mu_b]}$               & 65237                   & 187425                  & 217056                  & 196197                  \\
    $n_{[\nu_a,\nu_b]}$             & 872361                  & 608676                  & 333763                  & 173998                  \\
    $n_{[\nu_a,o_b]}$               & 5771                    & 17831                   & 21864                   & 15378                   \\
    $n_{[o_a,\nu_b]}$               & 5640                    & 16056                   & 20796                   & 18656                   \\
    $n_{[o_a,o_b]}$                 & 13                      & 260                     & 771                     & 1122                    \\
    $n_{[2\nu_a,2\nu_b]}$           & 65586                   & 23156                   & 35696                   & 17204                   \\
    $m_{[2\nu_a,2\nu_b]}$           & 16656                   & 6115                    & 10300                   & 5061                    \\
    $n_{[2\nu_a,o_b]}$              & 461735                  & 332943                  & 156652                  & 65098                   \\
    $n_{[o_a,2\nu_b]}$              & 409785                  & 267584                  & 151144                  & 85492                   \\ \midrule
    $E_{[\mu_a,\mu_b]}$             & 0.0007                  & 0.0022                  & 0.0079                  & 0.0128                  \\
    $E_{[2\nu_a,2\nu_b]}$           & 0.2540                  & 0.2641                  & 0.2885                  & 0.2941                  \\
    $\underline{n}_{11}^z$          & 39163294                & 29808789                & 13689739                & 5908436                 \\
    $\overline{e}_{11}^{\text{ph}}$ & 0.0717                  & 0.1312                  & 0.1124                  & 0.1511                  \\ \midrule
    SKR (bpp)                       & $1.7406 \times 10^{-5}$ & $1.1261 \times 10^{-6}$ & $2.2914 \times 10^{-7}$ & $2.0993 \times 10^{-8}$ \\
    SKR (bps)                       & 8695.34                 & 561.26                  & 113.59                  & 10.20                   \\
    PLOB                            & $1.1673 \times 10^{-4}$ & $1.0969 \times 10^{-6}$ & $9.0819 \times 10^{-8}$ & $9.9810 \times 10^{-9}$ \\
    SKR / PLOB                      & 0.1491                  & 1.0267                  & 2.5230                  & 2.1033                  \\ \bottomrule            
    \end{tabular}
    \label{tab_exp_results}
\end{table}

\clearpage

\bibliographystyle{spphys}
\bibliography{Ref}

\end{document}